\documentclass[%
aps,prl,twocolumn,
superscriptaddress,
]{revtex4-2}

\usepackage{graphicx}
\usepackage{amsfonts}
\usepackage{amssymb}
\usepackage{amsthm}
\usepackage{amsmath}
\usepackage{multirow}

\usepackage{color}
 \usepackage{makecell}
\usepackage{mathtools}

\usepackage[thinc]{esdiff}
\usepackage[colorlinks=true,allcolors=blue]{hyperref}

\begin{document}
	
\title{Magnetic monopoles and high frequency gravitational waves from quasi-stable strings}
\author{Rinku Maji}
\affiliation{Cosmology, Gravity and Astroparticle Physics Group, Center for Theoretical Physics of the Universe,  Institute for Basic Science, Daejeon 34126, Republic of Korea}
	\author{Qaisar Shafi}
		\affiliation{Bartol Research Institute, Department of Physics and 
		Astronomy,
		 University of Delaware, Newark, DE 19716, USA}
\begin{abstract}
The spontaneous  breaking of $SO(10)$ via flipped $SU(5)$  to the Standard Model yields a novel scenario in which the superheavy topologically stable GUT monopole carrying a single unit ($2\pi/e$) of Dirac magnetic charge emerges from the merger of a confined but topologically distinct monopole-antimonopole pair that are pulled together by a string. The $SO(10)$ breaking via the subgroup $SU(4)_c\times SU(2)_L\times SU(2)_R$, following a similar reasoning,  produces a topologically stable monopole that carries two units ($4\pi/e$) of Dirac charge. We explore the cosmological consequences of this scenario by assuming that the monopoles and strings experience a limited number of inflationary $e$-foldings, before re-entering the horizon and ultimately forming a network of quasi-stable strings bounded by monopole-antimonopole pairs. We identify regions of the parameter space that yield an observable number density of the GUT monopole from the collapse of the appropriate string segments. The gravitational waves emitted by these quasi-stable cosmic strings lie in the Hz to kHz range, which can be tested in a number of proposed and ongoing experiments.

\end{abstract}

\maketitle
\section{Introduction}

Dirac, nearly a century ago, postulated the existence of magnetic monopoles in order to explain electric charge quantization \cite{Dirac:1931kp}. Spontaneously broken gauge theories with quantized electric charge predict the existence of topologically stable magnetic monopoles \cite{tHooft:1974kcl,Polyakov:1974ek}. In the framework of grand unified theories such as $SU(5)$ \cite{Georgi:1974sy} and SO(10) \cite{Georgi:1974my,Fritzsch:1974nn}, the minimal monopole is superheavy and carries a single unit ($2 \pi/e$) of Dirac magnetic charge, as well as some color magnetic charge which is screened \cite{Daniel:1979yz, Dokos:1979vu, Lazarides:1982jq}. Unified theories based on a gauge symmetry such as $SU(4)_c \times SU(2)_L \times SU(2)_R$ \cite{Pati:1974yy} predict the existence of a magnetic monopole that carries two units ($4 \pi/e$) of Dirac charge as well as color magnetic charge \cite{Lazarides:1980cc, Lazarides:2019xai}. The mass of this monopole  can vary over a wide range, as discussed recently \cite{Lazarides:2024niy, Kephart:2025tik}.  For a discussion on how topologically stable monopoles can arise from the decay of a metastable string network, see Ref.~\cite{Lazarides:2024niy}.  A discussion of how superheavy monopole may survive primordial inflation can be found in Ref.~\cite{Moursy:2024hll}. For recent studies on metastable and superconducting strings in $SO(10)$ and $E_6$, see Refs.~\cite{Afzal:2023kqs,Maji:2024pll,Maji:2025thf,Maji:2025itv}. High frequency gravitational waves radiated from composite topological structures such as `walls bounded by strings' \cite{Kibble:1982dd} along with an observable flux of GUT monopoles, have been discussed in Ref.~\cite{Maji:2025yms}. For other related studies see Refs.~\cite{Buchmuller:2019gfy,Buchmuller:2021mbb,Dunsky:2021tih, Senjanovic:2025enc,Jeong:2025vbg, Moursy:2025ljr, Bandyopadhyay:2025oju, Bao:2025ori, Tranchedone:2026lav}.  For a discussion of the primordial monopole problem, see Refs.~\cite{Zeldovich:1978wj, Preskill:1979zi}.

In Ref.~\cite{Lazarides:2023iim} it was shown that in $SO(10)$, the GUT monopole can arise from the merger of an unrelated monopole-antimonopole pair, which get connected by a string carrying suitable magnetic flux. Our investigation is inspired by this observation and we propose to explore its consequences taking into account the inflationary scenario.  We assume that the primordial monopoles  experience a limited number of $e$-foldings and subsequently re-enter the horizon. Quantum tunneling of monopole-antimonopole pairs on the strings is exponentially suppressed, but the strings, referred to as quasi-stable strings, eventually disappear because they end up connecting monopole-antimonopole pairs \cite{Martin:1996ea, Lazarides:2022jgr}. A subset of these structures yield the topologically stable GUT monopoles. The quasi-stable strings also emit gravitational waves in the Hz to kHz range. We identify regions of the parameter space that yield an observable number density of the GUT monopole as well gravitational waves that can be measured in a variety of experiments.

 The detection of very high frequency gravitational waves (kHz to GHz) from quasi-stable strings  with an observable number density of topologically stable magnetic monopoles, as discussed here, can provide a unique probe of $SO(10)$ symmetry breaking, namely via flipped $SU(5)$ or $SU(4)_c\times SU(2)_L\times SU(2)_R$ subgroups. A wide variety of detector concepts (see Ref.~\cite{Aggarwal:2025noe} and the references therein) aim to probe such very high frequency regime of the gravitational wave spectrum.
 However, the proposed detector sensitivities still remain weaker \cite{Aggarwal:2025noe} than the integrated bound inferred from Big Bang Nucleosynthesis (BBN) and the cosmic microwave background (CMB) data \cite{Planck:2018vyg,EscuderoAbenza:2020cmq,Akita:2020szl,Froustey:2020mcq,Bennett:2020zkv}.  A stronger sensitivity of resonant electromagnetic cavities has been proposed in Refs.~\cite{Herman:2020wao,Herman:2022fau}.

\section{Monopoles connected by strings in SO(10) symmetry breakings}
\label{sec:2}

 The $SO(10)$ breaking via flipped $SU(5)$ that we consider is as follows:
\begin{align}
\label{eq:fsu5-chain}
SO(10) & \to SU(5)\times U(1)_X \nonumber \\ & \to SU(3)_c \times SU(2)_L \times U(1)_Z \times U(1)_X \nonumber \\ & \to SU(3)_c \times SU(2)_L \times U(1)_Y.
\end{align}
We normalize the $U(1)$ generators, following Ref.~\cite{Slansky:1981yr}, such that they have the minimal integer charges compatible with a period of $2\pi$. For example, we use the decompositions
\begin{align}
& SO(10) \to SU(5) \times U(1)_X: 
 16 = 10(-1) + \bar{5}(3) + 1(-5);
\nonumber \\
& SU(5)\to SU(3)_c \times SU(2)_L \times U(1)_Y: \nonumber \\
& 10 =  (3,2)(1) + (\bar{3}, 1)(-4) + (1,1)(6), \nonumber \\ & 
\overline{5} = (\overline{3},1)(2) + (1,2)(-3).
\end{align}
\begin{figure}[h!]
\begin{center}
\includegraphics[width=0.45\textwidth]{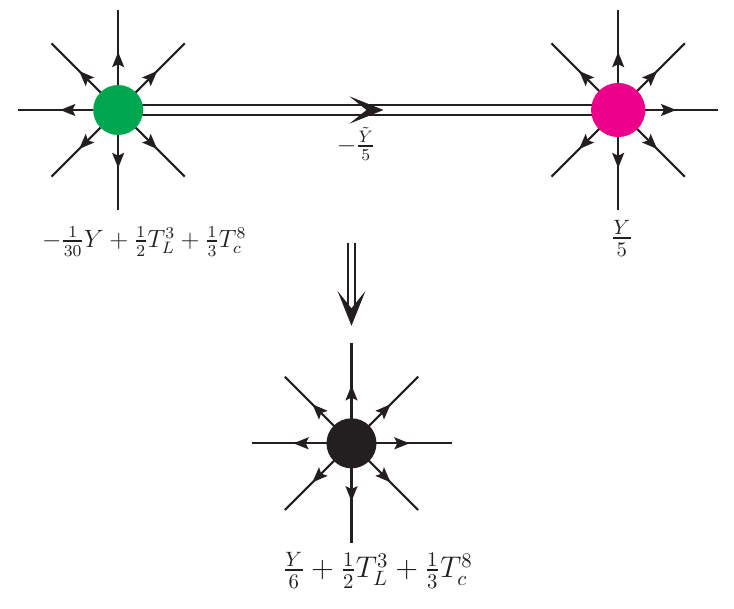}
\caption{Merger of $U(1)_X$ and $U(1)_Z$ monopoles before the electroweak symmetry breaking.}\label{fig:SO_fsu5_merger}
\end{center}
\end{figure}
\begin{figure}[h!]
\begin{center}
\includegraphics[width=0.45\textwidth]{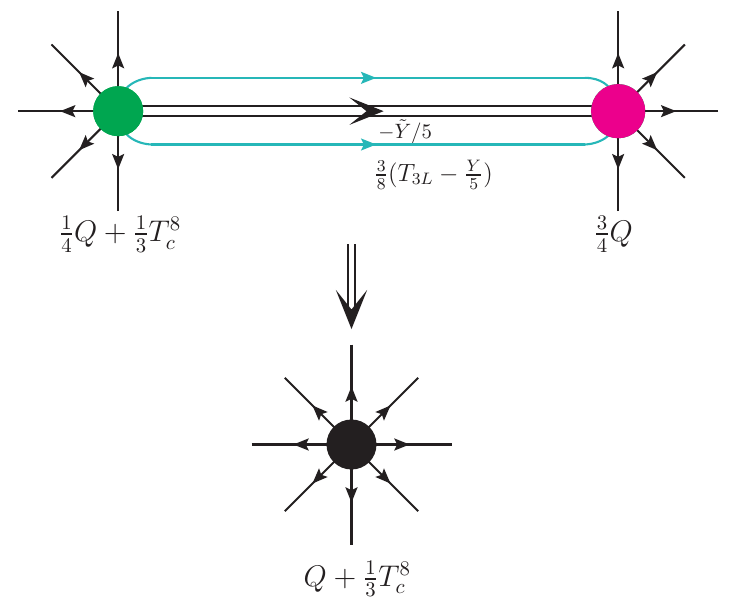}
\caption{ Emergence of $SO(10)$ GUT monopole with magnetic charge $2\pi/e$ from the merger of a $U(1)_X$ monopole and a $U(1)_Z$ antimonopole following the electroweak symmetry breaking.}\label{fig:SO_fsu5_merger_2}
\end{center}
\end{figure}
The first breaking produces `green' monopoles that carry $U(1)_X$ and $SU(5)$ fluxes. The second breaking produces `pink' monopoles carrying $U(1)_Z$ as well as $SU(2)_L$ and $SU(3)_c$ fluxes. A minimally charged `green' monopole carries the flux associated with $-\tfrac{X+Z}{5}$, and a `pink' monopole has fluxes associated with $\tfrac{Z}{6}+\tfrac{1}{2}T^3_L+\tfrac{1}{3}T^8_c$. Here $T^8_c = \mathrm{diag}(1,1,-2)$ is a diagonal generator of $SU(3)_c$, and $T^3_L=\mathrm{diag}(1,-1)$ is the diagonal generator of $SU(2)_L$. Following the last breaking of Eq.~\eqref{eq:fsu5-chain}, the unbroken and broken generators are given by \cite{Lazarides:2023iim, Maji:2025thf}
\begin{align}
\label{eq:BU-gentrs-fsu5}
Y = -\frac{Z+6X}{5}, \quad \tilde{Y} = \frac{-4Z+X}{5}.
\end{align}

Note that the $Z_5$ subgroups of $U(1)_Y$ and $U(1)_{\tilde{Y}}$ coincide. The breaking $U(1)_X\times U(1)_Z\to U(1)_Y$ produces a $U(1)_{\tilde{Y}}~Z_5$  flux tube that  connects a pink and a green monopole, as shown in Fig.~\ref{fig:SO_fsu5_merger}, as well as monopoles to their respective antimonopoles. Also note that $Y+\tilde{Y}=-(X+Z)$ and $Z/6=-({Y+6\tilde{Y}})/{30}$.
The merger of the green and pink monopoles produces  the topologically stable monopole. The broken generator orthogonal to the electric charge generator $Q=Y/6 + T^3_L/2$ is given by \cite{Maji:2025thf}
\begin{align}
\label{eq:brokenQ}
\mathcal{B}=T^3_L/2-Y/10.
\end{align}
After electroweak symmetry breaking there are unconfined fluxes since $-Y/30+T^3_L/2=Q/4+3\mathcal{B}/4$ as shown in Fig.~\ref{fig:SO_fsu5_merger_2}. The GUT monopole after electroweak symmetry breaking carries a single unit $(2\pi/e)$ of Dirac magnetic charge.
\begin{figure}[h!]
\begin{center}
\includegraphics[width=0.45\textwidth]{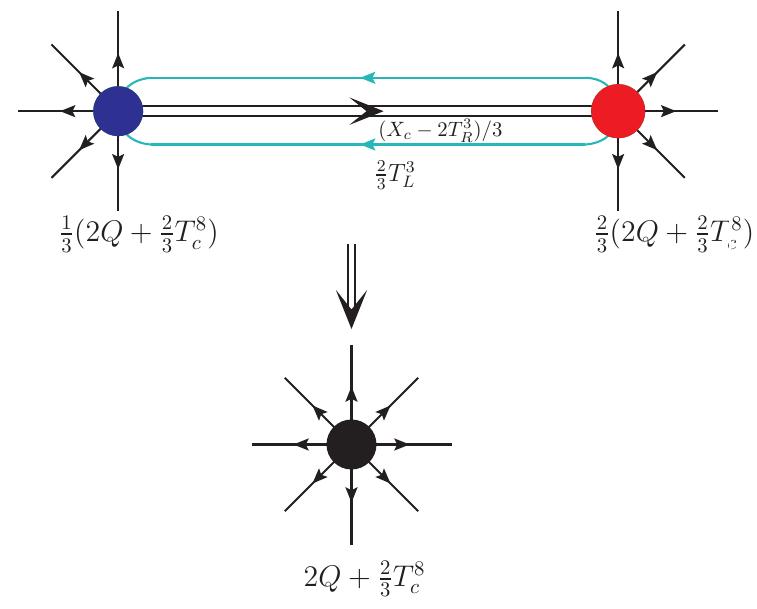}
\caption{Emergence of monopole of charge $4\pi/e$ from the merger of confined $U(1)_{R}$ and $U(1)_{B-L}$ monopoles following the electroweak symmetry breaking.} \label{fig:SO_PS_merger_2}
\end{center}
\end{figure}

For completeness, following Ref.~\cite{Lazarides:2019xai}, we briefly summarize here the appearance of a topologically stable monopole carrying two units of Dirac magnetic charge from the following $SO(10)$ breaking:
\begin{align}
\label{eq:422}
SO(10) &\to SU(4)_c \times SU(2)_L \times SU(2)_R \nonumber \\
&\to SU(3)_c \times SU(2)_L \times U(1)_{B-L} \times U(1)_R \nonumber \\ & \to SU(3)_c \times SU(2)_L \times U(1)_Y .
\end{align}
The second breaking produces a `blue' monopole carrying  $T^3_R$ flux, and a `red' monopole with flux associated with $X_c=(B-L)+\tfrac{2}{3}T^8_c$. During the third breaking a flux tube associated with a $2\pi/3$ rotation around the broken generator $(B-L)-2T^3_R$, orthogonal to $Y$, connects these two monopoles as shown in Fig.~\ref{fig:SO_PS_merger_2}. Of course, the monopoles can also merge with their respective antimonopoles.

 We discuss the simplest symmetry breaking patterns that produce the composite topological structures, namely monopole-antimonopoles connected by strings. The mergers of topologically distinct monopole-antimonopole pairs for the breaking patterns via flipped $SU(5)$ and $SU(4)_c\times SU(2)_L\times SU(2)_R$ produce topologically stable magnetic monopoles carrying a single unit  and two units of Dirac magnetic charge, respectively. As given in the examples, the symmetry breaking pattern in Eq.~\eqref{eq:fsu5-chain} is realized as the component $(1,0)$ and then $(24,0)$ in a $45$ dimensional Higgs acquire non-zero VEVs. On the other hand, the VEVs $(1,1,1)$ followed by $(15,1,3)$ from a 210 representation can realize the symmetry breaking pattern in Eq.~\eqref{eq:422}. In both cases, a VEV for the neutrino-like component in the $16$ dimensional scalar breaks the diagonal generator orthogonal to the SM hypercharge $Y$. It is worth mentioning here that introducing a $126$ dimensional scalar in place of $16$ leaves a $Z_2$ lying in the center of Spin(10) unbroken and leads to the formation of topologically stable cosmic strings \cite{Kibble:1982ae}. For recent studies on the dependence of the intermediate symmetry breaking scale on the choice of scalar representations, masses of scalar states and their phenomenological implications, see Refs.~\cite{Chakrabortty:2020otp, Preda:2022izo, Maji:2024tzg, Preda:2024vas, Preda:2025afo}. Note that the prediction of topological structures is independent of the supersymmetric or non-supersymmetric version of the GUTs. The unification scenarios may rely on the choice of Higgs representations (e.g. 45, 210, etc.), but this does not qualitatively alter the cosmological imprints such as gravitational waves and flux of monopoles discussed in the next section.

\section{Quasi-stable string and high frequency gravitational waves}
\label{sec:gws}
The confined monopoles experience partial $e$-foldings during inflation and re-enter the horizon at a time $t_M$. After horizon re-entry there will be subhorizon monopoles connected by strings carrying unconfined fluxes. Therefore, they decay dominantly by radiating the massless gauge bosons and generate topologically stable magnetic monopoles. The comoving number density of these monopoles can be estimated as \cite{Maji:2025yms}
\begin{align}
\label{eq:mon-yield-time}
Y_M\simeq\frac{1}{V_h s(t_M)}\simeq 
10^{-65}\frac{g_*(t_M)^{3/4}}{g_{*s}(t_M)}\left(\frac{\mathrm{sec}}{t_M}\right)^{3/2} ,
\end{align}
where $V_h = (2t_M)^3$ is the particle horizon volume during radiation domination. The upper bound on the flux of superheavy monopoles from MACRO experiment \cite{Ambrosio:2002qq} can be recast as $Y_M\lesssim 10^{-26}$ for monopole velocity $v_M\sim 10^{-3}$, which gives $t_M\gtrsim 10^{-26}$ sec. using Eq.~\eqref{eq:mon-yield-time}.

Before the horizon re-entry of the monopoles, the strings form a network of `stable' strings which we call quasi-stable string. The string network experiences friction domination until time $t_F\sim t_{\rm Pl}/(G\mu)^2$ \cite{Vilenkin:1991zk, Garriga:1993gj}, where $t_{\rm Pl}$ is the Planck time. If the horizon re-entry time $t_M$ of the monopoles is much later than the time scale of friction domination, the string network can enter the scaling regime at time $t_s$, which is typically two orders of magnitude higher than $t_F$ \cite{Martins:1995tg, Martins:1996jp, Martins:2000cs,Gouttenoire:2019kij}. The loops, formed after the domination of particle emission era $t_p\sim \tfrac{t_{\rm Pl}}{\Gamma^2 (G\mu)^{5/2}}$ \cite{Blanco-Pillado:1998tyu,Matsunami:2019fss, Auclair:2019jip}, can radiate gravitational waves.

 The present day ($t_0$) frequency of the gravitational waves in a normal mode $k$, radiated at time $\tilde{t}$ from a loop formed at time $t_i$ with an initial size $\alpha t_i$, is expressed as 
\begin{align}
f = \frac{a(\tilde{t})}{a(t_0)}\frac{2k}{\alpha t_i -\Gamma G\mu (\tilde{t}-t_i)} , \quad k\in Z^+,
\end{align}
where $a(t)$ denotes the scale factor, $\Gamma \simeq 50$, and $\alpha\simeq 0.1$ for a string network in the scaling regime \cite{Vachaspati:1984gt,Vilenkin:2000jqa}.
The gravitational wave background is given by
\begin{align}
\Omega_{\rm GW}(f) = \sum_{k=1}^{\infty}\Omega_{\rm GW}^{(k)}(f),
\end{align}
with
\begin{align}
   \label{eq:GWs2}
    \Omega^{(k)}_{\rm GW}(f) 
    = &\frac{1}{\rho_{c,0}} \int_{t_F}^{t_0} d\tilde{t} \left(\frac{a(\tilde{t})}{a(t_0)}\right)^5\frac{\mathcal{F} C_{\rm eff}(t_i)}{(\Gamma G \mu + \alpha)\alpha t_i^4} \left(\frac{a(t_i)}{a(\tilde{t})}\right)^3 \nonumber \\ & \frac{\Gamma k^{-n}}{\zeta(n)} G\mu^2 \frac{2 k}{f}\Theta(t_M - t_i) \Theta(t_i - \mathrm{max}[t_s,t_p]).
\end{align}
Here $n=4/3$ assuming the gravitational wave emission is dominated by the bursts from the cusp \cite{Damour:2001bk}, $\rho_{c,0}$ denotes the critical energy density of the universe at present, $\mathcal{F}\simeq 0.1$, and  $C_{\rm eff} = 5.7$ in the radiation dominated era \cite{Vanchurin:2005pa,Ringeval:2005kr,Olum:2006ix,Olmez:2010bi,Blanco-Pillado:2013qja,Blanco-Pillado:2017oxo,Cui:2018rwi}. 

\begin{figure}[ht]
\begin{center}
\includegraphics[width = 0.47\textwidth]{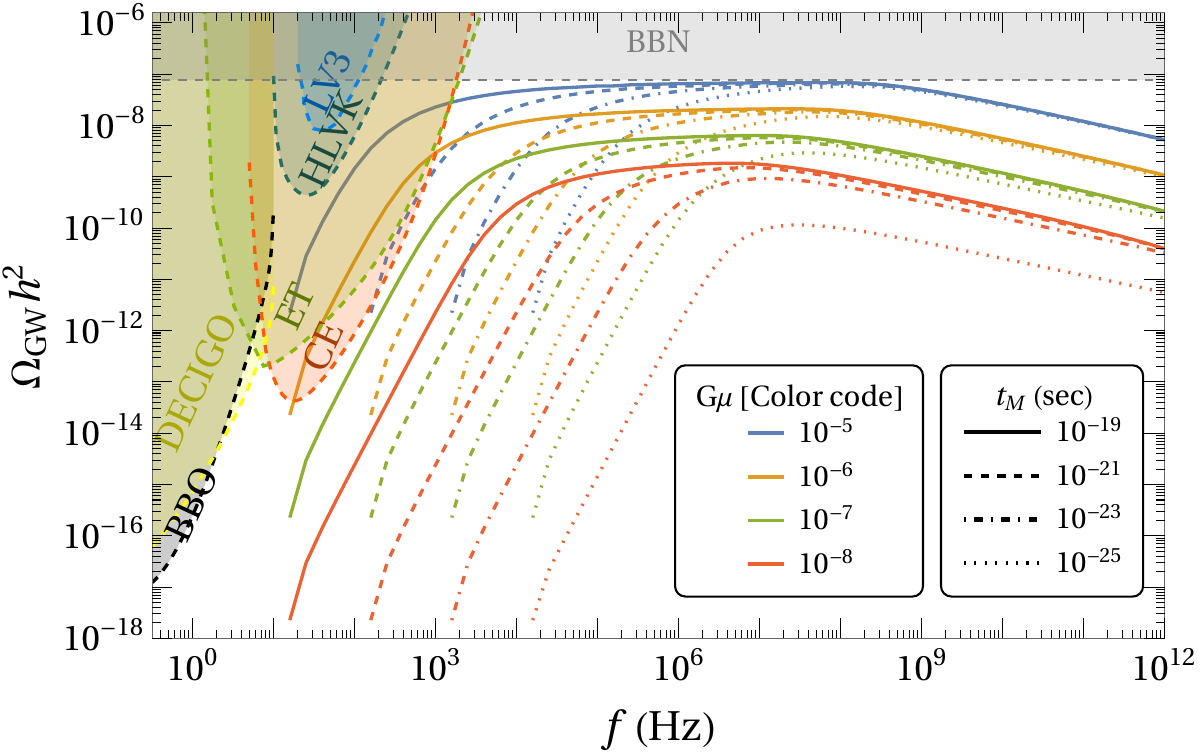}
\end{center}
\caption{Gravitational wave 
background from the quasi-stable string network for $G\mu = 10^{-8}-10^{-5}$ with a varying monopole horizon re-entry time $t_M=10^{-25}-10^{-19}$ sec, which can give rise to a comoving monopole number density $Y_M\sim 10^{-27}-10^{-37}$. We have depicted the power-law integrated sensitivity curves \cite{Thrane:2013oya, Schmitz:2020syl} for planned experiments near Hz to kHz frequency region, namely, HLVK \cite{KAGRA:2013rdx}, ET \cite{Mentasti:2020yyd},  CE \cite{Regimbau:2016ike}, DECIGO \cite{Sato_2017}, and BBO \cite{Crowder:2005nr, Corbin:2005ny}. The gray shaded region shows the bound for a scale-invariant gravitational wave spectrum for $f_{\rm high}/f_{\rm low}=10^7$ arising from the bound on $\Delta N_{\rm eff}$ in the BBN and CMB data.}
\label{fig:GWs_lowtM}
\end{figure}
At this stage, it is worth mentioning that the right-handed neutrinos achieve a Majorana mass $m_R\sim \mathcal{O}(v_s^2/\Lambda)$ ($\Lambda$ represents the UV cutoff scale) from the dimension five operators $\mathcal{C}_{\rm W} 16_F 16_H^\dagger 16_H^\dagger 16_F$ ($\mathcal{C}_{\rm W}$ represents the Wilsion coefficient), since the SM singlet in $16_H$ acquires a VEV $v_s$ \cite{Witten:1979nr, Preda:2022izo}. For $\Lambda\sim 10^{17}$ GeV and $\mathcal{C}_{\rm W}\sim 1$, we have $v_s\gtrsim 10^{15}$ GeV in order to obtain the heaviest right-handed neutrino mass $m_R\gtrsim 10^{12}$ GeV. The associated string has a dimensionless tension $G\mu\simeq \tfrac{1}{8}(v_s/m_{\rm Pl})^2\gtrsim 10^{-8}$, where $m_{\rm Pl}$ is the reduced Planck mass.

In Fig.~\ref{fig:GWs_lowtM} we display the gravitational wave background  for  $G\mu = 10^{-8}-10^{-5}$ with the monopoles horizon re-entry time $t_M=10^{-25}-10^{-19}$ sec, which can give rise to a comoving monopole number density $Y_M\sim 10^{-27}-10^{-37}$. The gravitational waves from GUT scale strings with $G\mu\sim 10^{-6}$ can be observed in the proposed experiments such as the Einstein Telescope (ET) \cite{Mentasti:2020yyd} and Cosmic Explorer (CE) \cite{Regimbau:2016ike} for $t_M\gtrsim 10^{-20}$ sec, along with a comoving monopole number density $Y_M\lesssim 10^{-36}$. The gravitational wave background with the monopole horizon re-entry time $t_M > 10^{-15}$ sec is ruled out by the third observing run data of LIGO and VIRGO \cite{LIGOScientific:2021nrg}. 

\begin{figure}[h!]
\begin{center}
\includegraphics[width = 0.47\textwidth]{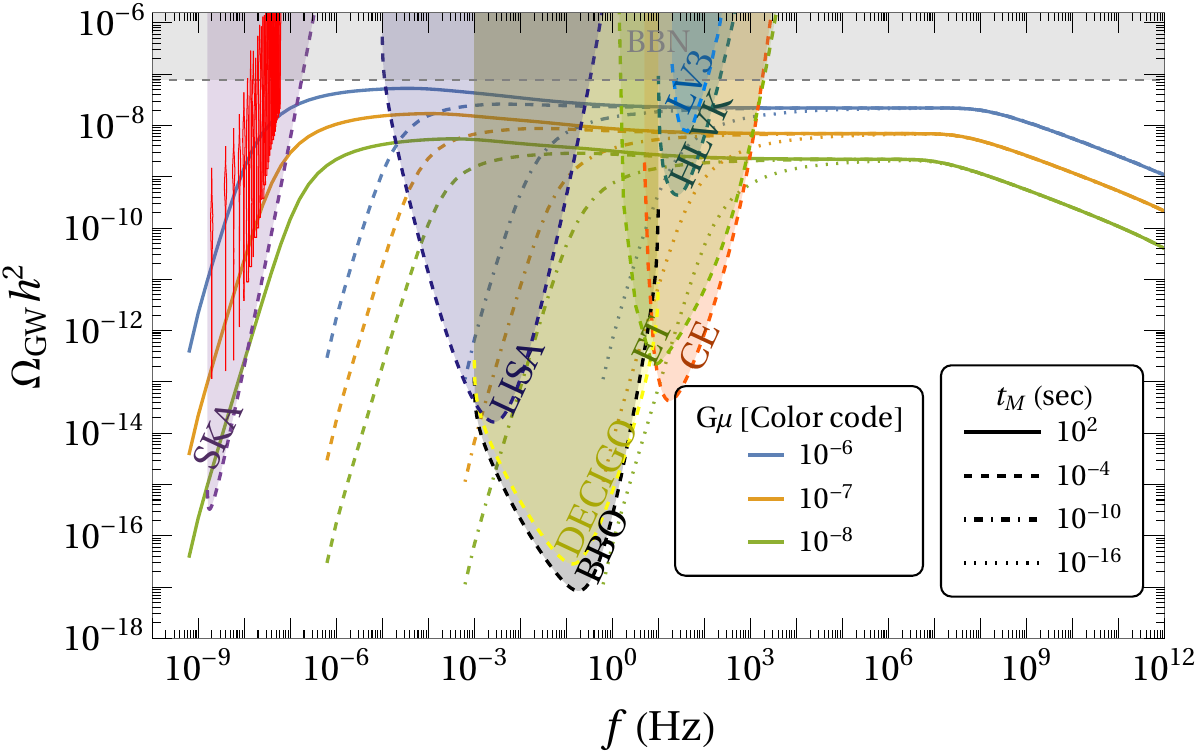}
\end{center}
\caption{Gravitational wave 
background from the quasi-stable string network for $G\mu = 10^{-8}$, $10^{-7}$, and $10^{-6}$ with a varying monopole horizon re-entry time $t_M=10^{-16}$, $10^{-10}$, $10^{-4}$ and $10^2$ sec. We also display the power-law integrated sensitivity curves \cite{Thrane:2013oya, Schmitz:2020syl} for proposed experiments, namely, HLVK \cite{KAGRA:2013rdx}, ET \cite{Mentasti:2020yyd},  CE \cite{Regimbau:2016ike}, DECIGO \cite{Sato_2017}, BBO \cite{Crowder:2005nr, Corbin:2005ny}, LISA \cite{Bartolo:2016ami, amaroseoane2017laser} and SKA \cite{Janssen:2014dka}. The gray shaded region shows the bound for a scale-invariant gravitational wave spectrum for $f_{\rm high}/f_{\rm low}=10^7$ arising from the bound on $\Delta N_{\rm eff}$ in the BBN and CMB data. The red violin plots represent the Bayesian free-spectral periodogram from the NANOGrav 15 year data \cite{NANOGrav:2023gor, NANOGrav:2023hvm}.}
\label{fig:GWs_hightM}
\end{figure}

The stochastic gravitational wave background from quasi-stable strings contributes to the effective number of relativistic degrees of freedom before big bang nucleosynthesis (BBN). The constraint on the gravitational wave spectra from BBN and cosmic microwave background (CMB) \cite{Aver:2015iza, Peimbert:2016bdg, Planck:2018vyg} can be expressed as~\cite{Maggiore:1999vm}:
\begin{align}\label{eq:bbn-cmb}
\frac{\rho_{\rm GW}}{\rho_c}=\int_{f_{\rm low}}^{f_{\rm high}}\Omega_{\rm GW}(f)d\ln{f} \lesssim \Omega_{\gamma,0} \frac{7}{8}\left(\frac{4}{11}\right)^{4/3}\Delta N_{\rm eff} ,
\end{align}
where the present day relic energy fraction in photons, $\Omega_{\gamma,0}h^2\simeq 2.5\times 10^{-5}$, and $\Delta N_{\rm eff}\lesssim 0.22$ is the combined upper bound from BBN and CMB data \cite{Planck:2018vyg,EscuderoAbenza:2020cmq,Akita:2020szl,Froustey:2020mcq,Bennett:2020zkv}. The lower limit on the frequency $f_{\rm low}$ is dictated by the gravitational radiation before BBN, whereas $f_{\rm high}$ is the highest frequency of the spectrum which could be taken as infinity without loss of the generality. Assuming a scale invariant spectrum $\Omega_{\rm GW}=\Omega_0$ for ${f_{\rm low}}\leq f \leq {f_{\rm high}}$, we can express the BBN bound as $\Omega_0h^2\lesssim 5.6\times 10^{-6}\Delta N_{\rm eff}/\log\left(\frac{f_{\rm high}}{f_{\rm low}}\right)$.

For completeness, we have depicted the gravitational wave spectra for $G\mu = 10^{-8},10^{-7}$, and $10^{-6}$ with the monopoles horizon re-entry time $t_M=10^{-16}$, $10^{-10}$, $10^{-4}$ and $10^2$ sec in Fig.~\ref{fig:GWs_hightM}. We have adopted $Y_M\gtrsim 10^{-36}$ as a threshold for observability. For $t_M>10^{-20}$ sec monopoles will be heavily diluted to be observed. 

It is worth mentioning at this point that the quasi-stable strings can explain the evidence of a gravitational wave background in the recent NANOGrav 15 year data \cite{NANOGrav:2023gor, NANOGrav:2023hvm} for $G\mu \in [2\times 10^{-8},7\times 10^{-6}]$ and $t_M\in [2\times 10^1, 1\times 10^5]$ sec (95\% Bayesian credible intervals) \cite{Lazarides:2023ksx, Maji:2024tzg}. The gravitational waves for $G\mu=[10^{-8},10^{-7}]$ can be detected in all the proposed experiments depending on the horizon re-entry time of the monopoles. The constraint from the third observing run data from LIGO and VIRGO experiments for $G\mu>10^{-7}$ can be alleviated if the strings experience a certain number of $e$-foldings during inflation for high $t_M$ values. For example, $G\mu =10^{-6}$ and $t_M=10^{2.5}$ sec can explain the NANOGrav data and satisfy the LV3 bound for $t_s=10^{-10}$ sec, as shown in Ref.~\cite{Maji:2024tzg}.


\section{Conclusions}
\label{sec:conc}
We have explored a novel symmetry breaking pattern of $SO(10)$  via its subgroup flipped $SU(5)$ which yields the topologically stable GUT monopole with a minimal Dirac charge of
$2\pi/e$ arising from the merger of a monopole with an unrelated antimonopole. The breaking of $SO(10)$ via the subgroup $SU(4)_c\times SU(2)_L\times SU(2)_R$ produces, on the other hand, an intermediate to superheavy scale monopole with a Dirac magnetic charge of $4\pi/e$, also from the merger of a topologically distinct pair of monopoles. In both cases, the strings responsible for these mergers form a quasi-stable string network and emit gravitational waves  over a wide range of frequencies with a scale-invariant spectrum upto GHz frequencies. We explore the parameter space that simultaneously yields observable gravitational waves and primordial magnetic monopoles.  We also find that the IR tail of the gravitational waves emitted by the superheavy quasi-stable cosmic strings can lie in the Hz to kHz range, which can be tested in a number of proposed experiments such as the Einstein Telescope and Cosmic Explorer. In the case with heavily diluted primordial monopoles, the predicted gravitational wave spectrum can explain the recently reported evidence by the pulsar timing array experiments for a gravitational wave background in the nHz frequency range.
\section{Acknowledgments}
RM is funded by IBS under the project code: IBS-R018-D3.
\appendix

\bibliographystyle{JHEP}
\bibliography{sccs}

\end{document}